\documentclass[journal=jctcce,manuscript=article]{achemso}

\usepackage[version=3]{mhchem} % Formula subscripts using \ce{}

\usepackage{bm}
\usepackage{graphicx}

\newcommand{\subscript}[1]{\ensuremath{_{\textrm{\footnotesize{#1}}}}}
\newcommand{\ket}[1]{\left| #1 \right\rangle}

\newcommand{\bra}[1]{\left\langle #1 \right|}

\author{J. P. Coe}
\email{J.Coe@hw.ac.uk}
\affiliation{Institute of Chemical Sciences, School of Engineering and Physical Sciences, Heriot-Watt University, Edinburgh, EH14 4AS, United Kingdom}

\title[\texttt{achemso} ]{Machine Learning Configuration Interaction}

\begin{document}
%%%%%%%%%%%%%%%%%%%%%%%%%%%%%%%%%%%%%%%%%%%%%%%%%%%%%%%%%%%%%%%%%%%%%
%% The manuscript does not need to include \maketitle, which is
%% executed automatically.  The document should begin with an
%% abstract, if appropriate.  If one is given and should not be, the
%% contents will be gobbled.
%%%%%%%%%%%%%%%%%%%%%%%%%%%%%%%%%%%%%%%%%%%%%%%%%%%%%%%%%%%%%%%%%%%%%
\begin{abstract}
We propose the concept of machine learning configuration interaction (MLCI) whereby an artificial neural network is trained on-the-fly to predict important new configurations in an iterative selected configuration interaction procedure. We demonstrate that the neural network can discriminate between important and unimportant configurations, that it has not been trained on, much better than by chance.  MLCI is then used to find compact wavefunctions for carbon monoxide at both stretched and equilibrium geometries. We also consider the multireference problem of the water molecule with elongated bonds. Results are contrasted with those from other ways of selecting configurations: first-order perturbation, random selection and Monte Carlo configuration interaction. Compared with these other serial calculations, this prototype MLCI is competitive in its accuracy, converges in significantly fewer iterations than the stochastic approaches, and requires less time for the higher-accuracy computations.
\end{abstract}


\section{Introduction}

Machine learning has become increasingly popular and successful as a tool for quantum chemistry, partly due to the advent of graphical processing unit training for deep neural networks.
Impressive applications have included training artificial neural networks, using energies from electronic structure calculations, to construct empirical potential energy surfaces for molecular dynamics, see, for example, Ref.~\citenum{Gastegger15} and references
therein. Another approach is to choose molecular descriptors as inputs then train the machine learning algorithm on density-functional theory (DFT) data to predict quantum chemical properties, for example spin-state gaps of transition-metal complexes.\cite{Janet17}
While a deep tensor neural network has been trained to accurately predict energies, when supplied with interatomic distances and nuclear charges, for molecules as large as salicylic acid.\cite{Schutt17}
The correlation energy has also been predicted using machine learning for large sets of organic molecules.\cite{Ramakrishnan15}  Furthermore, corrections to the energy of a variant of second-order M{\o}ller-Plesset perturbation theory have been successfully demonstrated 
using deep neural networks with inputs from quantum chemistry calculations.\cite{McGibbon17}

At a more fundamental level of quantum chemistry calculation, a deep neural network has been developed and trained to predict the Kohn-Sham kinetic energy within DFT when using the density as an input.\cite{Yao16} A powerful machine learning approach was later developed\cite{Brockherde17} to predict the density directly from the potential for DFT calculations thereby avoiding the need to solve the Kohn-Sham equations.  In Ref.~\citenum{Mills17}, machine learning was used to predict exact ground-state energies for one-electron systems using their two-dimensional potentials. There the deep neural network approach was found to be more efficient and could achieve chemical accuracy.

Much of machine learning for quantum chemistry uses DFT to generate the training data. Although DFT is exact in principle, in practice it depends on the choice of the approximate functional.  Elegant methods using coupled-cluster theory, CCSD\cite{CCSD} and CCSD(T),\cite{CCSD(T)} offer more reliable accuracy for systems that can be well-described by small corrections to a single Slater determinant. However their computational cost is greater than DFT and, in common with existing approximate functionals, they can perform very poorly when confronted with systems that have multiple important determinants. Such systems are often termed multireference problems and encompass stretched geometries, molecules containing transition metals, and excited states. The powerful approach of complete active space self-consistent field (CASSCF)\cite{siegbahn:2384} allows qualitatively correct results on multireference problems and can be followed by second-order perturbation (CASPT2)\cite{CASPT2} or truncated configuration interaction (MRCI)\cite{efficientMRCI} for quantitative accuracy. This, however, can come at a very high computational price and the results are now dependent on the choice of orbitals for the active space which can cause bias in, for example, the computation of potential energy surfaces.  

By capitalizing on the common paucity of configurations that contribute substantially to the full configuration wavefunction (FCI), selected configuration interaction methods iteratively construct a compact wavefunction that can approach the energy of FCI but do not require the user to choose an active space. If we could efficiently select only the important configurations then we could quickly converge to a compact and accurate wavefunction for multireference problems.  With this in mind, we take a different approach to machine learning in quantum chemistry by proposing an artificial neural network, trained on-the-fly, that predicts the important configurations for inclusion in a iterative selected configuration interaction calculation with the goal of improving accuracy and accelerating convergence.

Early work in selected configuration interaction (CI) included using first-order perturbation theory for the coefficients of the wavefunction to select new configurations (CIPSI),\cite{CIPSI} or the contribution to the energy from perturbation theory to second order.\cite{HarrisonFCIperturbation} Later, the Monte Carlo configuration interaction (MCCI) method was developed\cite{dissociationGreer,mcciGreer98,mccicodeGreer} which stochastically adds configurations to build up the wavefunction where configurations that are found to have absolute coefficients less than the control parameter (c\subscript{min}) are removed. 

Recently there has been a resurgence of interest in selected CI. For example, MCCI has been built upon and successfully applied to other challenges in quantum chemistry including crossings of potential curves for excited states,\cite{2013saMCCI} molecular tunnel junctions formed of gold atoms,\cite{GreerComplexMCCI} hyperpolarizabilities,\cite{MCCIhyper} perturbative corrections for the dissociation of diatomics,\cite{MCCIfirstrowdiss}  X-ray absorption in multireference molecules,\cite{XrayMCCI15} and spin-orbit coupling.\cite{MCCIspinorbit}
The $\Lambda$-CI method, developed in Ref.~\citenum{LambdaCI}, chooses configurations that are within a prescribed energy from the lowest energy configuration and was demonstrated to successfully calculate dissociation curves for N\subscript{2} and the carbon dimer.
Adaptive configuration interaction (ACI) was later created\cite{adaptiveCI} which uses both energy and coefficients as criteria to select configurations and was shown to give good agreement with DMRG results for singlet-triplet splittings
in acenes. ACI was then built upon to give accurate results for excited states.\cite{adaptiveCIexcited} Adaptive sampling CI (ASCI)\cite{deterministicFCIQMC} improves upon CIPSI by only generating single and double substitutions from configurations with the largest coefficients. Instead of using the full expression of first-order perturbation to select configurations as in CIPSI, heat-bath CI\cite{HeatBathCI} employs a simpler quantity and its impressive results are in agreement with DMRG for the chromium dimer.  The MCI3 method was created in Ref.~\citenum{MC3I} by using projector or diffusion Monte Carlo in configuration space\cite{ProjectorMCNagase08,AlaviFCIQMC1} to approximate the first-order perturbative corrections to the wavefunction. These are then used to guide the selection of new configurations thereby allowing larger systems to be considered than for CIPSI.  MCI3 was applied to the  ground and excited states of the carbon dimer where it gave accurate potential curves when compared with FCI but used a very small fraction of the configurations. CIPSI has also been recently used to create trial wavefunctions for diffusion Monte Carlo calculations that give the correct ground state of FeS.\cite{CIPSIforDMC}

In this paper we first discuss the methods employed beginning with the artificial neural network, then the MCCI program which is used as the framework for the other selected-CI approaches. This leads in to the description of machine learning configuration interaction (MLCI) which uses the artificial neural network trained on-the-fly to select important configurations.  We also describe replacing the neural network predictions with predictions from first-order perturbation or random prediction. The accuracy of the neural network predictions are then investigated on stretched carbon monoxide in the 3-21G basis and this multireference system is then used to compare final energies, iterations to convergence and timings for the four selected-CI approaches. 
We then consider the molecule at its equilibrium bond length to assess the prototype MLCI method on a problem that is not multireference. Finally, we check that the form of the neural network can also work well for selected-CI calculations on other molecules by applying the methods
to the water molecule with a cc-pVDZ basis and stretched bonds as an example of another multireference problem.

\section{Methods}

\subsection{Artificial Neural Network}

\begin{figure}[h!]\centering
\includegraphics[width=.6\textwidth]{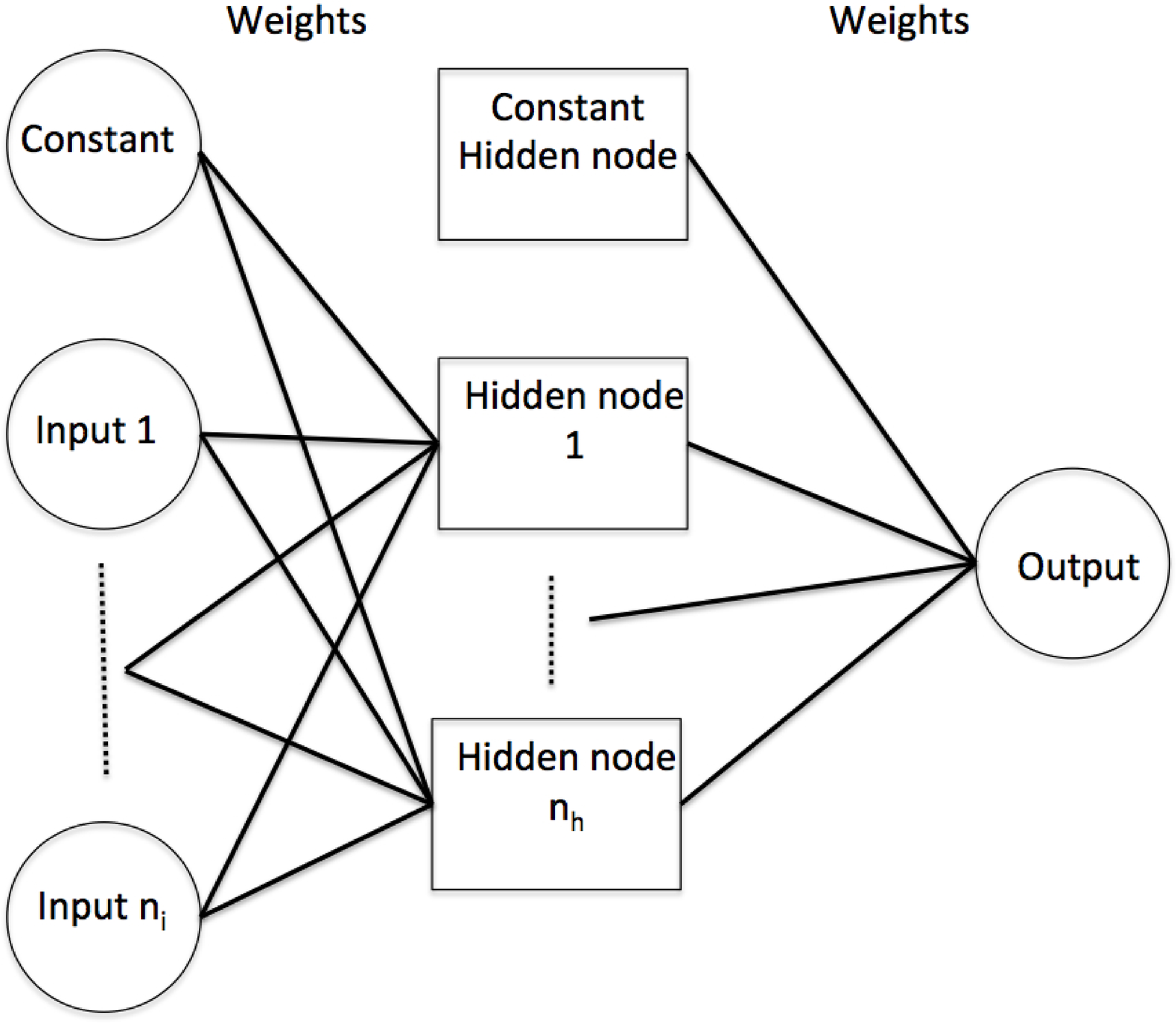}
\caption{Schematic of the artificial neural network with $n_{i}$ inputs plus a constant input, a single layer of $n_{h}$ hidden nodes plus a constant hidden node, and one output.}\label{neuralSchematic}
\end{figure}

The general form of the artificial neural network used is depicted in Fig.~\ref{neuralSchematic} where we have $n_{i}$ inputs plus a constant input, a single layer of $n_{h}$ hidden nodes plus a constant hidden node, and one output. Hidden nodes are just nodes that are neither an input nor an output. 
For a basis set of size $M$ we use $2M$ inputs, corresponding to the spin orbitals, and one constant input for the neural network.  An input is $1$ if that spin orbital is occupied in the configuration of interest and $0$ otherwise. 
After trialling various numbers of hidden nodes we settled on $30$ hidden nodes plus one constant. 

 We label the weights from input $i$ to hidden node $j$ as $W^{in}_{ij}$ while
those from the hidden layer $j$ to the output are labelled $W^{out}_{j0}$.
The value of hidden node $j$ is given on $(0,1)$ by a logistic or sigmoid function:
\begin{equation}
\text{Hidden}_{j}=\frac{1}{1+e^{-\sum_{i=0}^{n_{i}} W^{in}_{ij} \text{Input}_{i}}},
\end{equation}
where $i$ runs from $0$ to $n_{i}$ as input node $0$ is constant.  The output is between zero and one, and is calculated by 
\begin{equation}
\text{Output}=\frac{1}{1+e^{-\sum_{j=0}^{n_{h}} W^{out}_{j0} \text{Hidden}_{j}}}.
\end{equation}

Without a hidden layer, then finding the weights for a given set of input and output values would just be logistic regression which could be transformed to linear regression when the training output is on $(0,1)$. However a single hidden layer, in principle,
allows the neural network to approximate essentially any function when sufficient nodes are used.\cite{1HiddenLayerSufficient}  Many of the recent successful applications of neural networks have employed multiple hidden layers which are known as deep or convoluted neural networks.
It is not clear whether this is because deep networks are less dependent on the form of the input or are easier to train with current techniques than using a single hidden layer with many nodes. Interesting work has shown that shallow neural networks can give similar accuracy
when trained using the outputs of previously trained deep neural networks.\cite{DoNetsNeedToBeDeep}  For this proof-of-concept work we restrict the neural network to a single hidden layer.

We train the neural network to approximate the outputs of a training set of input and output values $o_{t}$ using backpropagation with stochastic gradient descent, see, for example, Ref.~\citenum{MitchellBook97}.
The weights are initially set to small random values then for each training set example the output is calculated and compared with the desired result $o_{t}$.  Stochastic gradient descent is used to minimize the error 
\begin{equation}
\text{error}=\frac{1}{2}(\text{output}- o_{t})^{2}  
\end{equation}
through updating the weights for every training example error rather than using gradient descent on the total error for the whole training set. By using this approach the chance of being trapped in a local minimum is hoped to be reduced as it is tantamount to introducing noise into gradient descent.  
When the errors are propagated back to update the weights, the change in the weights on each iteration is controlled by the learning rate parameter.  This balances speed of training against likely accuracy.  We use an initial learning rate of 0.1 that drops to 0.01 from the third iteration of machine learning configuration interaction.

  Sufficient training of the neural network can require many passes through the entire training set.  However the neural network needs to perform well on unseen data not just the training set. So, to avoid very high quality results on the training set but very poor quality predictions for new data (overtraining), the neural network is applied to a verification set of values that it has not been trained on at each iteration. One technique is then to stop training the neural network once its error on the verification set begins to increase.  For this work, the maximum number of passes through a training set is fixed at 2000 and we use the weights that give the lowest error on the verification set from these 2000 training passes.

 We implement an artificial neural network for selected configuration interaction within the framework of the Monte Carlo configuration interaction program,\cite{dissociationGreer,mcciGreer98,mccicodeGreer} which we discuss next.

\subsection{Monte Carlo Configuration Interaction}
Monte Carlo Configuration Interaction (MCCI)\cite{dissociationGreer,mcciGreer98,mccicodeGreer} builds up a compact wavefunction by repeatedly randomly adding interacting configurations to the current set then diagonalizing the Hamiltonian matrix and eventually removing configurations 
whose absolute coefficient is less than the cutoff c\subscript{min}. 

For MCCI in this work we begin with the Hartree-Fock Slater determinant and a brief description of the algorithm is presented below:
\begin{itemize}
\item{Symmetry-preserving single and double substitutions are used to stochastically enlarge the current configuration space.}
\item{The Hamiltonian matrix is constructed and diagonalized to give the energy and coefficients of the wavefunction.}
\item{If a configuration is newly added but has a coefficient less than c\subscript{min} then it is deleted (pruned).}
\item{All configurations in the wavefunction are considered for deletion (full pruning) every ten iterations.}
\item{This process is iterated until the energy after a full pruning step satisfies a convergence criterion.\cite{GreerMcciSpectra}}
\end{itemize}

The convergence threshold\cite{GreerMcciSpectra} is set to c\subscript{min} here and uses the energy change between full pruning steps averaged over the last three full pruning steps and compares the maximum of the last three values of this with the threshold.  
For all the work in this paper we use Slater determinants as the configurations and run the calculations in serial. The Hartree-Fock molecular orbitals are calculated using Molpro,\cite{Molpro_2015} as are the one-electron and two-electron integrals.

\subsection{Machine Learning Configuration Interaction}

To facilitate the training of the neural network to predict important configurations in machine learning configuration interaction (MLCI) we transform the coefficients $c_{i}$ of configurations to $|\tilde {c}_{i}|$ for the neural network.  With the aim of making the difference between important and unimportant configurations more apparent.  We set absolute coefficients
on $[0,c\subscript{min})$ to zero and linearly map absolute coefficients on $[c\subscript{min},1]$ to $[0.6,1]$. This latter transformation is achieved using
\begin{equation}
|\tilde{c}_{i}| = \frac{0.4|c_{i}|+0.6-c\subscript{min}}{1-c\subscript{min}}.
\end{equation} 
We note that if the weights are all zero then the logistic function will give a value of 0.5 so we required a value greater than this and found that 0.6 as the threshold was satisfactory.  These $|\tilde {c}_{i}|$ and their configurations are respectively the outputs and inputs for the training set.

The neural network requires initial data for training so the first step in MLCI is to add all single and double substitutions permitted by spatial symmetry to the Hartree-Fock wavefunction, i.e., we create the CISD wavefunction. 
 
Pruning is then implemented to remove new configurations in the wavefunction with $|c_{i}|<c\subscript{min}$. As all information on the importance of configurations is useful for training and we do not want the neural network to `forget' what an unimportant configuration looks like, we store pruned configurations as a reject set without duplicates and with zero coefficients. If a pruned configuration later becomes important it is removed from this set.  That the neural network does not reproduce the training data perfectly may help the calculation: the training data contains a snapshot of the importance of configurations for the current wavefunction and so is an evolving approximation as the importance may change when configurations are included and removed.  Hence just because a configuration is in the reject set at one point in the calculation does not mean it should never be added again.  

The transformed coefficients and their configurations in both the wavefunction and the reject set are then shared randomly and equally between the training set and the verification set. The neural network is then trained with 2000 passes through the training set and the weights that give the lowest error on the verification set are used for the next step. 

The neural network is then applied to all single and double substitutions from the current wavefunction and, for a wavefunction of $L$ configurations, adds the $L$ configurations it considers to be most important, i.e., they have the largest
predicted coefficients. As the number of single and double substitutions from the current set can become very large we use the method of Ref.~\citenum{MCCInatorb} to efficiently generate all single and double substitutions without duplicates. We note that the neural network's weights are random at the start of MLCI but then are retained between iterations.
The neural network, in a sense, pulls itself up by its bootstraps as after the first iteration in MLCI it both predicts the configurations to be added and learns, on-the-fly, from the output of the diagonalizations.

The MLCI procedure is summarized below where initially we construct the CISD wavefunction from the Hartree-Fock single determinant.
\begin{itemize}
\item{Create and diagonalize the Hamiltonian matrix in the current configuration space.}
\item{Newly added configurations with $|c_{i}|<c\subscript{min}$ are pruned and stored in the reject set. Every ten iterations the entire wavefunction is pruned (full prune).}
\item{The neural network is trained to predict transformed coefficients $|\tilde{c}_{i}|$ by combining the reject set with the wavefunction and sharing these data equally and randomly between training and verification sets.}
\item{All symmetry-allowed single and double substitutions without duplicates are efficiently generated\cite{MCCInatorb} from the pruned wavefunction.}
\item{For a wavefunction of $L$ configurations the neural network then predicts the best $L$ new configurations, i.e. those with the largest predicted $|\tilde{c}_{i}|$, to enlarge the current configuration space.}
\item{This procedure is iterated until the energy satisfies a convergence criterion.\cite{GreerMcciSpectra}}
\end{itemize}

We note that MLCI reaches convergence much sooner than MCCI so in this work MLCI looks at every iteration, not just every ten, to check for convergence. This ensures that the MLCI procedure is not repeated unnecessarily many times.

\subsection{Prediction By First-Order Perturbation}
We also consider first-order perturbation theory to select configurations, which we term PT prediction. This therefore uses the same approach for selection as CIPSI.\cite{CIPSI}
The procedure is that of MLCI except instead of using the neural network for the predictions we use the absolute coefficients $|c_{I}|$ in the first-order correction $\Psi^{(1)}$ to the current wavefunction $\Psi^{(0)}$.
In this case
\begin{equation} 
\Psi^{(1)}=\sum_{I} c_{I} \ket{I}
\end{equation}
and
\begin{equation}
c_{I}=\frac{\bra{I} \hat{H} \ket{\Psi^{(0)}} } {E^{(0)} - \bra{I} \hat{H} \ket{I}},
\end{equation}
where the $\ket{I}$ are all the symmetry-allowed single and double substitutions when duplicates are excluded. For real coefficients then the wavefunction to first-order is $\Psi^{(0)}+\Psi^{(1)}$ which we normalize.  We then add the $L$ configurations with the largest absolute $c_{I}$ values to the current configuration space.  We found that the calculation of the $c_{I}$ is somewhat slower than applying the neural network in this work, but convergence is rapid so again we check for convergence on every iteration.

\subsection{Random Prediction}
To check that MLCI is doing better than random selection we finally replace the predicted values in the algorithm with random numbers on $[0,1]$.  This random prediction is not the same as MCCI as it is created by modifying MLCI so generates the CISD wavefunction on the first iteration, and creates all single and double substituted configurations from the current wavefunction without duplicates then randomly selects $L$ of these configurations. In contrast, MCCI uses random generation on the first iteration and attempts to add an adaptive number of configurations where the probability of single or double substitution is the same.\cite{mccicodeGreer}  Hence random prediction enables a fairer comparison between the MLCI program and stochastic selection. As the convergence when using random prediction is rather slow then the convergence check only looks at full pruning steps as in MCCI, but unlike MLCI and PT prediction.

\section{Results}
\subsection{Stretched Carbon Monoxide}

We initially test the MLCI approach on carbon monoxide with a stretched bond of $4$ Bohr, the 3-21G basis set and two frozen orbitals. This system, with four frozen orbitals, was shown to be strongly multireference in Ref.~\citenum{MRinQC} when using the canonical Hartree-Fock orbitals. 
A cutoff of $c$\subscript{min}$=10^{-3}$ is used within the MCCI framework when we first consider the ability of the neural network to predict important configurations. Where, as discussed in the Methods section, absolute coefficients on $[0,c_{\text{min}})$ are set to zero
and absolute coefficients on $[c_{\text{min}},1]$ are linearly transformed to $[0.6,1]$ to improve the training of the neural network in discerning important configurations. 

\begin{figure}[h!]\centering
\includegraphics[width=.6\textwidth]{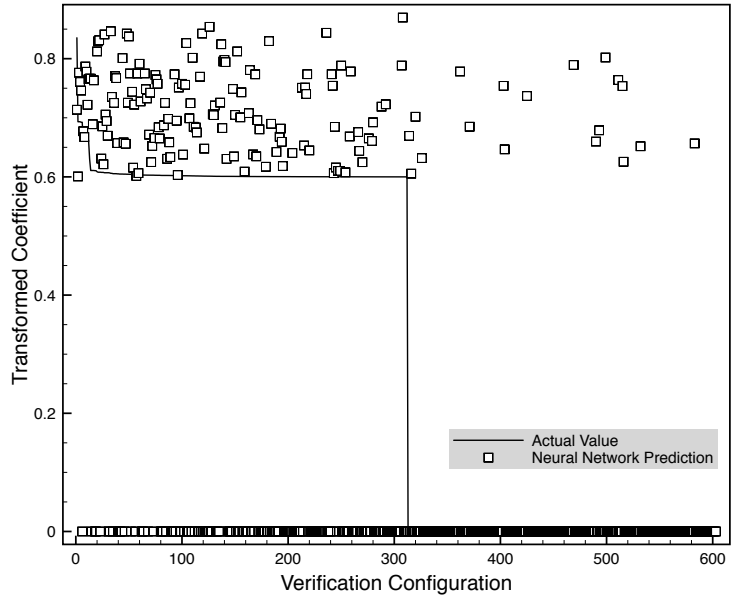}
\caption{Transformed coefficients $|\tilde{c}_{i}|$ of the verification set configurations compared with the neural network predictions, when values below $0.6$ are set to zero, for carbon monoxide using a bond length of $4$ Bohr, the 3-21G basis
set, two frozen orbitals and a cutoff of c\subscript{min}$=10^{-3}$ on the first iteration in MLCI.}\label{stretchCOverify1}
\end{figure}

The neural network results from the first iteration of MLCI are presented in Fig.~\ref{stretchCOverify1} where we order the configurations in the verification set by the size of their transformed coefficients and compare these coefficients with the neural network predictions when values less than 0.6 are set to zero. At this early stage of the computation, the
neural network improved its accuracy by running through the training set 772 times before its error on the verification set began to increase. The training set consisted of 603 configurations, and the verification set was the same size.   At this point the root-mean-square error on the verification set was $0.27$. The neural network can be seen (Fig.~\ref{stretchCOverify1}) to be predicting the importance of the configurations fairly well although a number of important configurations are not identified and a few unimportant configurations are classified incorrectly.

The predictive ability of the neural network on the verification set from the first iteration of MLCI is quantified in Table~\ref{tbl:stretchCOverify1}. There we see that, at this point, the ability of the neural network to include important configurations has room for improvement as using 
\begin{equation}
\text{sensitivity}=\frac{\text{true positives}}{\text{true positives+false negatives}}
\end{equation}
then the sensitivity, or true positive rate, is $46\%$.  Yet we emphasize that the MLCI algorithm limits the number of configurations to be added using the size of the current wavefunction, so it does not necessarily matter if we miss some important configurations. At this early stage of developing MLCI we would be satisfied to find significantly more important configurations than would be expected by chance. Particularly as this advantage will accumulate over multiple iterations.  With regards to this the neural network is much better at not including unimportant configurations as using
\begin{equation}
\text{specificity}=\frac{\text{true negatives}}{\text{true negatives+false positives}}
\end{equation}
then the specificity, or true negative rate, is $94\%$.  We note that of the 162 configurations suggested to be included by the neural network, 144 ($89\%$) are actually important.  Around half of the verification set are important in the wavefunction so one would expect that if the 162 configurations were instead chosen randomly then only about 80 would be important. 

\begin{table}[h!]
\centering
\caption{Neural network importance predictions compared with whether the configuration is important in the verification set's transformed coefficients $|\tilde{c}_{i}|$  for the first iteration of MLCI for carbon monoxide using a bond length of $4$ Bohr, the 3-21G basis set, two frozen orbitals, a cutoff of c\subscript{min}$=10^{-3}$ and when values less than 0.6 are considered unimportant.}
\begin{tabular}{@{\extracolsep{\fill}}lcc}
\hline
 	 & Important  & Unimportant  \\
\hline
Predicted Important  &     144 & 18 \\
Predicted Unimportant &   168 & 273 \\
\hline
\label{tbl:stretchCOverify1}
\end{tabular}
\end{table}

\begin{figure}[h!]\centering
\includegraphics[width=.6\textwidth]{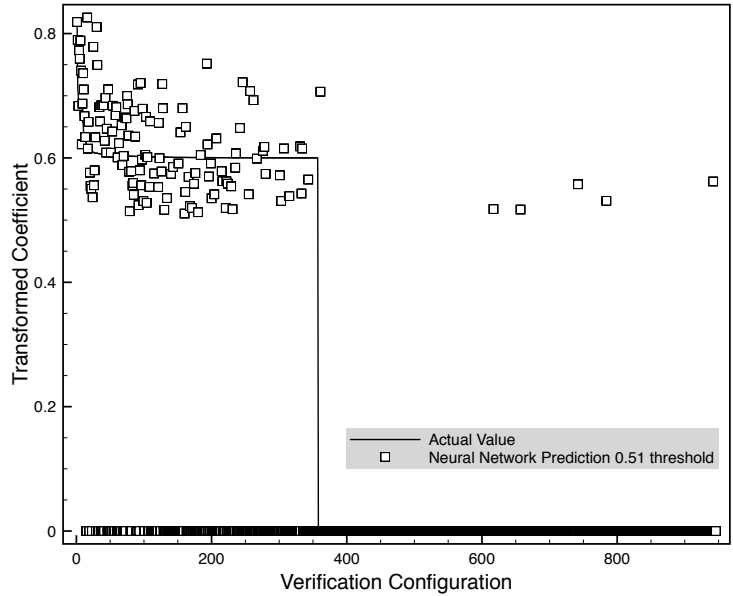}
\caption{Transformed coefficients $|\tilde{c}_{i}|$ of the verification set configurations compared with the neural network predictions with a threshold of $0.51$ for carbon monoxide using a bond length of $4$ Bohr, the 3-21G basis
set, two frozen orbitals and a cutoff of c\subscript{min}$=10^{-3}$ on the second iteration in MLCI.}\label{stretchCOverify2}
\end{figure}

We now plot the neural network predictions of the second iteration in Fig.~\ref{stretchCOverify2}.  In this case we use $0.51$ rather than $0.6$ as the threshold below which neural network predictions are mapped to zero in the figure.  The MLCI algorithm includes the $L$ configurations with 
the predicted largest coefficients where $L$ is the number of configurations in the current wavefunction. Hence it is the relative importance of the predictions that matters and we can see from 
Fig.~\ref{stretchCOverify2} that with the lower threshold there are many true positives with only a small number of false positives. However it can also be seen that there are
many configurations that are important but not classified as such. For these results the neural network weights were taken after 89 passes through the training set as after this the verification error began to increase. The root-mean-square error on the 
verification set is now $0.20$ hence the neural network has continued to improve from the first iteration of MLCI despite the increase in the number of configurations in the verification set.

This change in accuracy on lowering the threshold is quantified in Table~\ref{tbl:stretchCOverify2} where we see that the number of false positives only increases from $1$ to $5$ but the number of true positive
goes up from $72$ to $129$ as the threshold drops to $0.51$. In this case, the sensitivity increases from $20.2\%$ to $36.1\%$ while the specificity only slightly decreases from $99.8\%$ to $99.0\%$. For the lower threshold, 135 configurations would be predicted to be important and $95.6\%$ are indeed important in the current wavefunction. In the second iteration for MLCI, $37.9\%$ of the verification set were important in the wavefunction so by picking 135 configurations randomly one would only expect around 51 to be important.

\begin{table}[h!]
\centering
\caption{Neural network importance predictions compared with whether the configuration is important in the verification set's transformed coefficients $|\tilde{c}_{i}|$  on the second iteration of MLCI, when values less than 0.6 are considered unimportant for the verification set and 0.6 then 0.51 for the neural network. Results are for carbon monoxide using a bond length of $4$ Bohr, the 3-21G basis
set, two frozen orbitals and a cutoff of c\subscript{min}$=10^{-3}$.}
\begin{tabular}{@{\extracolsep{\fill}}lcc}
\hline
 	 & Important  & Unimportant  \\
\hline
Predicted Important ($\geq 0.6$) &     72 & 1 \\
Predicted Unimportant ($<0.6$) &   285 & 588 \\
Predicted Important  ($\geq 0.51$) &     129 & 6 \\
Predicted Unimportant ($<0.51$) &   228 & 583 \\
\hline
\label{tbl:stretchCOverify2}
\end{tabular}
\end{table}

\subsubsection{Energy Calculations}
\begin{figure}[h!]\centering
\includegraphics[width=.6\textwidth]{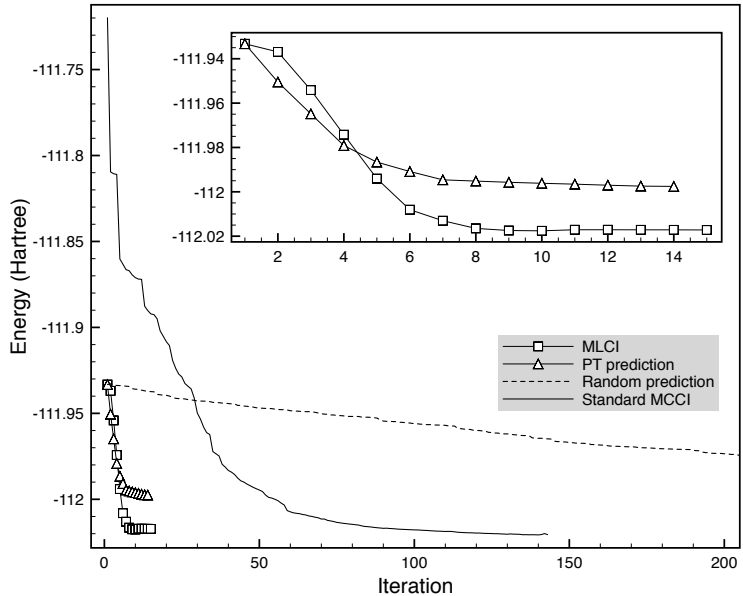}
\caption{Energy (Hartree) against iteration number for serial calculations for carbon monoxide with a bond length of $4$ Bohr, the 3-21G basis
set, two frozen orbitals and a cutoff of c\subscript{min}$=10^{-3}$ using machine learning predictions to select configurations (MLCI),  PT prediction, random prediction or standard MCCI. Inset: Enlargement of the MLCI and PT prediction curves.}\label{stretchCOconvergence}
\end{figure}

We have seen that the neural network can perform well on verification sets, but the crucial test is its ability, when confronted with the much larger set of single and double substitutions, to accelerate the convergence of a selected configuration interaction calculation.
Fig.~\ref{stretchCOconvergence} shows that MLCI converges in around one tenth of the iterations required for MCCI when $c$\subscript{min}$=10^{-3}$ and the convergence tolerance for the energy is $10^{-3}$ Hartree. We note that all calculations are run in serial and by running MCCI in parallel then convergence would require fewer iterations.

Although the MLCI convergence required one more iteration than using PT prediction, the final energy
is noticeably lower using MLCI (inset of Fig.~\ref{stretchCOconvergence}).  As MCCI does not include all single and doubles on the first iteration then for a direct comparison of machine learning versus random additions we also modify MLCI so that the predictions from
the neural network are replaced by random numbers.  This random prediction approach is also presented in Fig.~\ref{stretchCOconvergence} and we can see that, although the energy is similar at the start due to the addition of all singles and doubles
on the first iteration, the convergence is very much slower than MLCI and also of MCCI.  It could be that building up the wavefunction more slowly at the beginning is advantageous when randomly adding configurations. Perhaps as this reduces the chance of configurations that will eventually become unimportant being in the wavefunction and so the singles and doubles space can be smaller and also more likely to have important configurations.   We note that MCCI has other differences with random prediction which could contribute to its faster convergence rather than just the number of configurations added on the first iteration. For example, there is an equal chance of adding a single or double substituted configuration in MCCI while in the singles and doubles space the doubles are much more numerous.

\begin{figure}[h!]\centering
\includegraphics[width=.6\textwidth]{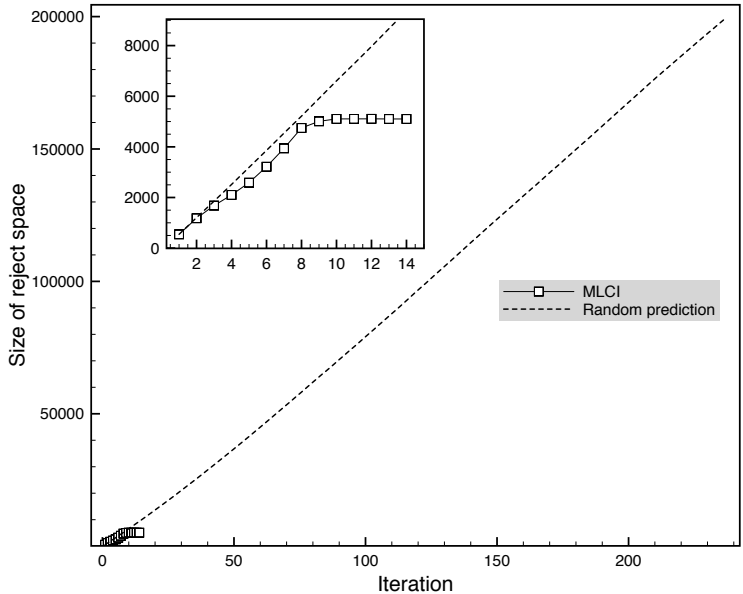}
\caption{Size of the reject space against iteration number for serial calculations for carbon monoxide with a bond length of $4$ Bohr, the 3-21G basis
set, two frozen orbitals and a cutoff of c\subscript{min}$=10^{-3}$ using machine learning predictions to select configurations (MLCI) or random prediction. Inset: Enlargement of the results until convergence of MLCI.}\label{stretchCOrejects}
\end{figure}

 As we construct the three predictive approaches to always add the same number of configurations as in the current wavefunction, it is not the case that MLCI adds more configurations than PT or random prediction on an iteration. Rather the configurations
suggested by the neural network are more likely to turn out to be important in the wavefunction.  We see in Fig.~\ref{stretchCOrejects} that predicting important configurations randomly means that the size of the reject space
increases almost linearly with the number of iterations as very many added configurations are pruned. In fact this causes the calculation to end as the size of the wavefunction and the size of the reject space
are limited to $2\times10^{5}$ configurations. MLCI, in contrast, has fewer of its predicted configurations pruned from the wavefunction on each iteration and as it converges the size of the reject space changes very little as the neural network will be predicting essentially the same configurations on each iteration if both the wavefunction and neural network have converged.

We emphasize that these are proof-of-concept calculations for predicting important configurations so that the algorithms and code could be made more efficient, but we present indicative timing information in Table \ref{tbl:stretchCOtiming}.  The results for random prediction here differ to those plotted in Figs.~\ref{stretchCOconvergence} and \ref{stretchCOrejects} in that they limit the number of rejects to the length of the wavefunction so that the number of rejects does not cause the calculation to stop and convergence can be reached.  The time required
for MLCI is not much longer than MCCI despite MLCI generating all the singles and doubles without duplicates. This also slows down random prediction and PT prediction, where in addition the former suffers from the large number of iterations with little energy change while the latter has the cost of evaluating $\bra{I} \hat{H} \ket{\Psi^{(0)}}$ for all members of the singles and doubles space. This means that they take noticeably more time to converge than the other two approaches. PT prediction also calculates the normalization for the first-order wavefunction
which is not strictly necessary if we are only interested in the relative magnitude of coefficients in this wavefunction and will contribute to the time cost of the calculation. The FCI energy was calculated with Molpro\cite{Molpro_2015} and required $\sim4.8$ million Slater determinants.  The MCCI wavefunction
used the most configurations, but also captured the most correlation energy while MLCI was not far behind. Using a multireference indicator (MR) for configuration interaction wavefunctions\cite{MCCImetaldimers,MRinQC} we find a value of $0.94$ for the MLCI wavefunction compared with that of $0.95$ for MCCI.  MR approaches one as the wavefunction becomes very strongly multireference so this system is indeed a strongly multireference problem and MLCI is capturing the multireference character.

\begin{table}[h!]
\centering
\caption{Percentage of the FCI correlation energy recovered, iterations, time and number of configurations for converged serial calculations using MLCI, random prediction, PT prediction or MCCI for carbon monoxide with a bond length of $4$ Bohr, the 3-21G basis
set, two frozen orbitals and c\subscript{min}$=10^{-3}$.}
\begin{tabular}{@{\extracolsep{\fill}}lcccc}
\hline
 	 & $\%$ Correlation Energy  & Iterations & Time (Seconds) & Configurations \\
\hline
MLCI     &  93.9    & 15  &	325	& 2477\\
Random Prediction &   87.7 & 333 &	1763 &	1897\\
PT Prediction  &    88.0 & 14	&   4349     &	1924\\
MCCI &   95.3 & 143 &	291	& 2853 \\
\hline
\label{tbl:stretchCOtiming}
\end{tabular}
\end{table}

In Table \ref{tbl:stretchCOtimingLowerCmin} we see that, when the cutoff is lowered to c\subscript{min}$=5\times10^{-4}$, the MLCI calculation is around twice as fast
as the MCCI example run although MLCI captures slightly less of the correlation energy using fewer configurations. Again, very few iterations are needed for the convergence of MLCI or PT prediction and the MLCI energy is closer to
the FCI result than using random or PT predictions. Similar to the c\subscript{min}$=10^{-3}$ result, we see that random prediction requires very many iterations for convergence although its final energy is now more accurate than using PT prediction.   Table \ref{tbl:stretchCOtimingLowerCmin} also shows that if we  lower the cutoff further to c\subscript{min}$=2\times10^{-4}$ then MLCI is around 3.5 times faster than MCCI. However, again, slightly less of the correlation energy is
recovered by MLCI in its 16 iterations although this is not at odds with the fewer configurations that it requires to give $98.3\%$ of the correlation energy for this multireference problem.

\begin{table}[h!]
\centering
\caption{Percentage of the FCI correlation energy recovered, iterations, time and number of configurations for converged serial calculations using MLCI, random prediction, PT prediction or MCCI for carbon monoxide with a bond length of $4$ Bohr, the 3-21G basis
set, two frozen orbitals and c\subscript{min}$=5\times10^{-4}$ or c\subscript{min}$=2\times10^{-4}$.}
\begin{tabular}{@{\extracolsep{\fill}}lcccc}
\hline
 	 & $\%$ Correlation Energy  & Iterations & Time (Seconds) & Configurations \\
\hline
c\subscript{min}$=5\times10^{-4}$ \\
MLCI     & 96.9	&15&	703&	5638\\
Random Prediction & 94.7 &	523&	7714&	5146   \\
PT Prediction  & 91.4 & 17 &	16947	& 4114  \\
MCCI & 97.7 &	143&	1431&	6451    \\
\hline
c\subscript{min}$=2\times10^{-4}$ \\
MLCI     & 98.3 &	16	& 2079 &	12971\\
MCCI &   99.1 &	133	&7293 &	16014 \\
\hline
\label{tbl:stretchCOtimingLowerCmin}
\end{tabular}
\end{table}

\subsection{Equilibrium Carbon Monoxide}

We now consider carbon monoxide using, again, the 3-21G basis set and two frozen orbitals, but at its equilibrium bond length\cite{COdipoleExperiment} of $2.1316$ Bohr.  With four frozen orbitals such a system was previously found\cite{MRinQC} to not have significant multireference when using the canonical Hartree-Fock orbitals. For problems that are not multireference then one would expect approaches based on small perturbative corrections to be more efficient when starting from the Hartree-Fock determinant and PT prediction should be more accurate. 

\begin{figure}[h!]\centering
\includegraphics[width=.6\textwidth]{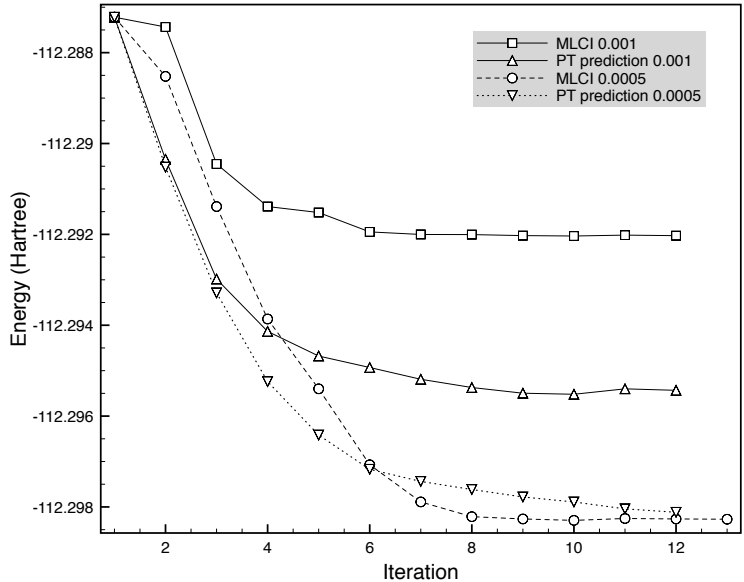}
\caption{Energy (Hartree) against iteration number for serial calculations for carbon monoxide with a bond length of $2.1316$ Bohr, the 3-21G basis
set, two frozen orbitals and a cutoff of c\subscript{min}$=10^{-3}$ or c\subscript{min}$=5\times 10^{-4}$ with c\subscript{min} also the convergence tolerance, using machine learning predictions to select configurations (MLCI) or PT prediction.}\label{eqCOconvergence}
\end{figure}

In Fig.~\ref{eqCOconvergence} the convergence in the energy is displayed only for MLCI and PT prediction, as the number of iterations were so much lower for these methods than the others.  For c\subscript{min}$=10^{-3}$ it is indeed the case that PT prediction gives a lower converged energy than MLCI. The energy scale of the graph amplifies this difference to a degree as PT prediction captures $93.7\%$ of the correlation energy while the MLCI is just a little lower at $92.3\%$.  Interestingly, on lowering the cutoff to c\subscript{min}$=5\times 10^{-4}$ we see in Fig.~\ref{eqCOconvergence} that
MLCI gives a very slightly lower energy than PT prediction, although PT prediction required one fewer iteration.

Table \ref{tbl:eqCOtiming0.0005} shows that for c\subscript{min}$=5\times 10^{-4}$ the MCCI run gave the closest energy to FCI and required a similar time to MLCI, which was the quickest approach here. Again the number of iterations required for convergence
were substantially more for the stochastic methods than for MLCI or PT prediction and, in this case, random prediction gave the highest energy.  A multireference indicator (MR) gives $0.16$ for both MCCI and MLCI suggesting that the wavefunction would not be considered multireference.
Despite this, the neural network in MLCI performs similarly to PT prediction in the accuracy of the energy (although PT prediction used slightly fewer configurations) and is only a little less accurate than MCCI which required noticeably more configurations.

\begin{table}[h!]
\centering
\caption{Percentage of the FCI correlation energy recovered, iterations, time and number of configurations for converged serial calculations using MLCI, Random prediction, PT prediction and standard MCCI for carbon monoxide with a bond length of $2.1316$ Bohr, the 3-21G basis
set, two frozen orbitals and c\subscript{min}$=5\times10^{-4}$.}
\begin{tabular}{@{\extracolsep{\fill}}lcccc}
\hline
 	 & $\%$ Correlation Energy  & Iterations & Time (Seconds) & Configurations \\
\hline
MLCI    & 95.2 &	13	&255	&2366  \\
Random Prediction & 92.9 &	103	&653	&1995  \\
PT Prediction  & 95.1 &	12	&3927&	2278 \\
MCCI &  96.9 &	113	&290	& 3350   \\
\hline
\label{tbl:eqCOtiming0.0005}
\end{tabular}
\end{table}

\subsection{Stretched Water}

To demonstrate that the form of neural network used can tackle not only the particular case of carbon monoxide (10 electrons and 16 orbitals) but also other multireference problems, we finally look at water with stretched bonds.
We use the cc-pVDZ basis with one frozen orbital resulting in 8 electrons in 23 orbitals.  The bond length is set to $4.8$ Bohr and the angle to $104.5$ degrees.

\begin{figure}[h!]\centering
\includegraphics[width=.6\textwidth]{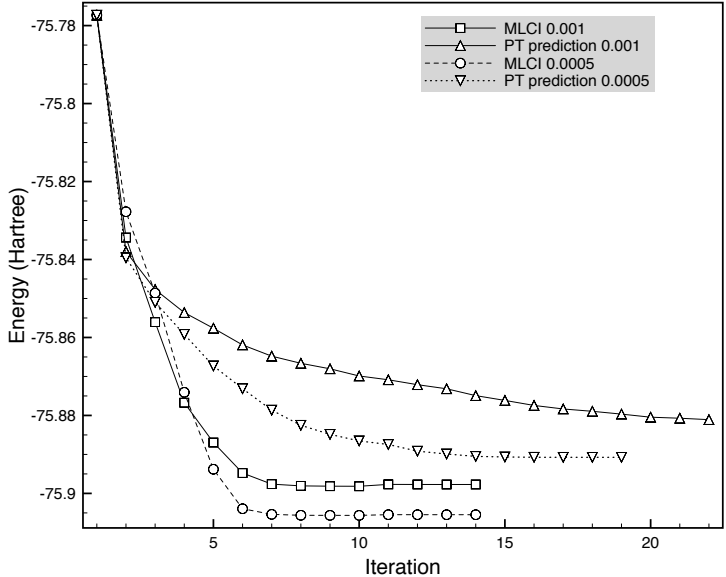}
\caption{Energy (Hartree) against iteration number for serial calculations for water with bond length $4.8$ Bohr, angle $104.5$ degrees, the cc-pVDZ basis
set, one frozen orbital and a cutoff of c\subscript{min}$=10^{-3}$ or c\subscript{min}$=5\times 10^{-4}$ with the same values for the convergence tolerance using machine learning predictions to select configurations (MLCI) or PT prediction.}\label{stretchH2Oconvergence}
\end{figure}

Fig.~\ref{stretchH2Oconvergence} displays the energy against iteration for MLCI and PT prediction as the other approaches required many more iterations. For this system
we see that MLCI gives a lower energy at convergence than PT prediction. In addition, the MLCI result at the larger cutoff (c\subscript{min}$=10^{-3}$) is actually slightly more accurate than PT prediction at the lower cutoff (c\subscript{min}$=5\times 10^{-4}$).

We see in Table \ref{tbl:stretchH2Otiming} that, for c\subscript{min}$=10^{-3}$, MLCI and MCCI both capture the most correlation energy to one decimal place but we mention that to two decimal places MLCI recovers marginally more  ($96.23\%$ versus $96.20\%$)
and uses slightly fewer configurations at 2086. However in this case the MLCI calculation was slower than MCCI.  Random prediction performs poorly due to a much earlier convergence than in the runs on the previous systems.

On lowering the cutoff to c\subscript{min}$=5\times10^{-4}$ we find that this is a strongly multireference problem as MCCI gives an MR value of $0.94$ while for MLCI the value is lower, but still indicative of strong multireference character, at $0.83$. However the MLCI value for multireference character could be the more accurate: when running MCCI with c\subscript{min}$=10^{-4}$ we found a wavefunction of 23259 configurations and an MR value of $0.84$. Returning to c\subscript{min}$=5\times10^{-4}$ and Table \ref{tbl:stretchH2Otiming}, we see that random prediction gives an energy that is not so different to the other approaches, but in doing so requires many more iterations than the other
methods. MLCI, now the quickest approach, captures only slightly less of the correlation energy than MCCI using fewer configurations: $3967$ versus $4368$ to give around $98\%$ of the correlation energy
where we note that the FCI wavefunction for this system comprises around 19.6 million configurations.

\begin{table}[H]
\centering
\caption{Percentage of the FCI correlation energy recovered, iterations, time and number of configurations for converged serial calculations using MLCI, Random prediction, PT prediction and MCCI for water with bond length $4.8$ Bohr, angle $104.5$ degrees, the cc-pVDZ basis set, one frozen orbital and c\subscript{min}$=10^{-3}$ or c\subscript{min}$=5\times10^{-4}$.}
\begin{tabular}{@{\extracolsep{\fill}}lcccc}
\hline
 	 & $\%$ Correlation Energy  & Iterations & Time (Seconds) & Configurations \\
\hline
c\subscript{min}$=10^{-3}$ \\
MLCI     & 96.2 &	14	&425	&2086 \\
Random Prediction & 73.4 &	63&	320 &	893    \\
PT Prediction  & 93.1 &	22&	14468&	1791  \\
MCCI & 96.2 &	193	&283&	2231    \\
\hline
c\subscript{min}$=5\times10^{-4}$ \\
MLCI     & 98.0 &	14&	686&	3967 \\
Random Prediction & 94.1 &	713 &	9469 &	3328 \\
PT Prediction  & 95.1 &	19 &	35526 &	3558 \\
MCCI &  98.3 &	173 &	  917&	4368  \\
\hline
\label{tbl:stretchH2Otiming}
\end{tabular}
\end{table}

\section{Summary}

In this paper we put forward the idea of machine learning configuration interaction (MLCI) for quantum chemistry. Here an artificial neural network is trained on-the-fly to select important configurations as it iteratively builds up a wavefunction by choosing configurations for inclusion in a selected configuration interaction scheme. 

For stretched carbon monoxide, we demonstrated how the chosen form of neural network could discriminate between important and unimportant configurations, that it was not trained on, much better than by chance. The MLCI procedure was applied to this multireference problem and shown to converge in significantly fewer iterations than when predicting important configurations randomly. We found Monte Carlo configuration interaction (MCCI)\cite{dissociationGreer,mcciGreer98,mccicodeGreer} to give the most accurate results for this system for a given cutoff (c\subscript{min}) which defines the minimum coefficient a configuration needs in the wavefunction to not be eventually deleted.  However MLCI was only a little less accurate, used fewer configurations and, even for the proof-of-concept MLCI program, required significantly less time for serial computation as c\subscript{min} decreased. However, we noted that the number of iterations to convergence for MCCI could be reduced if it was run in parallel.   Using first-order perturbation theory (PT prediction) to predict configurations instead of the neural network also resulted in significantly fewer iterations to convergence than stochastic approaches. However compared with the neural network approach much more computation time was necessary and, perhaps in keeping with this being a multireference problem, the energy was less accurate than MLCI. 

We then considered carbon monoxide at its equilibrium geometry as an example of a system that is not significantly multireference and should be well-described by methods built around small corrections to a single determinant. For larger c\subscript{min}, corresponding to less accurate calculations, we saw that PT prediction captured a little more of the correlation energy than MLCI. Interestingly when c\subscript{min} was lowered, the more accurate calculations had MLCI and PT prediction giving similar energies
and MLCI was noticeably faster. Again MCCI required many more iterations, but gave a slightly more accurate energy using more configurations than MLCI.

To verify that the form of the neural network was not just appropriate for carbon monoxide, we finally investigated the MLCI approach on the multireference problem of water with stretched bonds.
For c\subscript{min}$=5\times 10^{-4}$ we found that fewer than twenty iterations were needed for convergence of MLCI or PT prediction while the MCCI run required 173.  When replacing the neural network predictions
in MLCI with random predictions 713 iterations were necessary. MLCI, again the fastest method, captured only slightly less of the correlation energy ($98\%$) than MCCI and more than the other approaches. To do this it used around 4000 configurations compared
with approximately 19.6 million in the full configuration interaction wavefunction.

We have seen that machine learning configuration interaction (MLCI) can use on-the-fly training of an artificial neural network to iteratively build up a compact wavefunction that can capture much of the accuracy of full configuration interaction but using a very small fraction of the configurations. Despite MLCI being implemented as a prototype at this stage, we found that compared with the other ways considered here of selecting configurations its accuracy was competitive, it could converge in fewer than twenty iterations and it required less time for the higher-accuracy serial calculations on small molecules including systems with, and without, significant multireference character.

Although on-the-fly training of the artificial neural network did not disadvantage the relative speed of the calculations here, for larger basis sets and more configurations this could become a bottleneck. Future work will investigate using graphical processing units for training deep neural networks to enable MLCI to efficiently and accurately calculate ab initio potential energy surfaces of larger molecules.

%% classes.  Rather than use \section, an appropriate macro is
%% provided that will always work.
%%%%%%%%%%%%%%%%%%%%%%%%%%%%%%%%%%%%%%%%%%%%%%%%%%%%%%%%%%%%%%%%%%%%%
\begin{acknowledgement}
JPC thanks the EPSRC for support via the platform grant EP/P001459/1.
\end{acknowledgement}

%% \suppinfo macro.
%%%%%%%%%%%%%%%%%%%%%%%%%%%%%%%%%%%%%%%%%%%%%%%%%%%%%%%%%%%%%%%%%%%%%

%\mciteErrorOnUnknownfalse
\providecommand{\noopsort}[1]{}\providecommand{\singleletter}[1]{#1}%
\providecommand{\latin}[1]{#1}
\makeatletter
\providecommand{\doi}
  {\begingroup\let\do\@makeother\dospecials
  \catcode`\{=1 \catcode`\}=2\doi@aux}
\providecommand{\doi@aux}[1]{\endgroup\texttt{#1}}
\makeatother
\providecommand*\mcitethebibliography{\thebibliography}
\csname @ifundefined\endcsname{endmcitethebibliography}
  {\let\endmcitethebibliography\endthebibliography}{}

\end{document}